\documentclass[11pt,a4paper]{article}
\usepackage{amsmath,amssymb}
\usepackage{graphicx}
\usepackage{booktabs}
\usepackage{setspace}
\usepackage{geometry}
\usepackage{caption}
\usepackage{natbib}
\usepackage{doi}
\usepackage{url}
\usepackage{hyperref}
\usepackage{authblk}
\usepackage[utf8]{inputenc}
\usepackage{tikz}
\usetikzlibrary{positioning,arrows.meta,shapes.geometric}

\title{\textbf{Time-Varying Hazard Patterns and Co-Mutation Profiles of KRAS G12C and G12D in Real-World NSCLC}}

\author[1]{Robert Amevor\thanks{Corresponding author: ramevor@email.sc.edu}}
\author[2]{Emmanuel Kubuafor}
\author[2]{Dennis Baidoo}

\affil[1]{Arnold School of Public Health, University of South Carolina,
          Columbia, SC, USA}
\affil[2]{Department of Mathematics and Statistics, University of New Mexico,
          Albuquerque, NM, USA}

\date{}

\begin{document}
\maketitle

% ============================================================
\begin{abstract}

\noindent\textbf{Background:}
KRAS mutations represent the largest oncogenic subset in non--small cell lung cancer
(NSCLC), yet allele-specific clinical behavior in real-world settings remains
incompletely characterized. While KRAS G12C is now targetable with approved
inhibitors, no equivalent agents exist for G12D. We examined time-to-next-treatment
(TTNT) and overall survival (OS) differences between KRAS G12C and G12D using
harmonized clinical--genomic data, allowing for time-varying hazard effects.

\noindent\textbf{Methods:}
De-identified data from the AACR Project GENIE BioPharma Collaborative (BPC)
NSCLC v2.0-public release were analyzed. TTNT served as a pragmatic real-world
surrogate for progression-free survival, derived from curated systemic therapy
timelines. Co-alterations in \textit{TP53}, \textit{STK11}, \textit{KEAP1},
\textit{SMARCA4}, and \textit{MET} were harmonized across mutation, copy-number,
and structural variant data. Kaplan--Meier, multivariable Cox, time-dependent Cox,
and a pre-specified piecewise Cox model (split at median TTNT = 23.0 months) were
applied. Proportional hazards were assessed via Schoenfeld residuals; nonparametric
bootstrap resampling ($B = 1000$) evaluated stability of time-varying estimates.

\noindent\textbf{Results:}
The TTNT cohort comprised 162 patients (G12C $n=130$; G12D $n=32$). Median TTNT was
28.6 months for G12C versus 32.0 months for G12D (log-rank $p = 0.79$). Adjusted
Cox regression showed no overall TTNT hazard difference (HR$_{\text{G12D vs G12C}}
= 0.85$; 95\%~CI 0.53--1.37; $p = 0.50$). Schoenfeld testing indicated borderline
non-proportionality for KRAS ($p = 0.053$). Piecewise Cox modeling revealed a
time-varying pattern: an early TTNT hazard favoring G12D
(HR$_{\text{early}} = 0.41$; 95\%~CI 0.17--0.97; $p = 0.043$) with a statistically
significant KRAS$\times$period interaction (HR = 3.33; 95\%~CI 1.19--9.31;
$p = 0.021$) and attenuation in the late period
(HR$_{\text{late}} = 1.38$; 95\%~CI 0.77--2.47; $p = 0.285$). Bootstrap
resampling confirmed these patterns (median HR$_{\text{early}} = 0.39$;
median HR$_{\text{late}} = 1.41$). Among 278 OS-evaluable patients (133 deaths),
multivariable Cox regression indicated improved OS for G12D
(adjusted HR = 0.63; 95\%~CI 0.39--0.99; $p = 0.048$). KRAS G12C tumors exhibited
higher median TMB (9.79 vs. 7.83 mut/Mb; $p = 0.002$) and greater enrichment of
\textit{STK11} and \textit{KEAP1} co-mutations.

\noindent\textbf{Conclusions:}
KRAS G12D and G12C demonstrate distinct temporal treatment trajectories in this
multi-institutional real-world cohort. G12D was associated with an early TTNT
advantage and improved overall survival; these patterns were directionally consistent
across bootstrap and sensitivity analyses. The late-period TTNT difference did not
reach statistical significance, likely reflecting limited power (post-hoc power:
12.3\%) given the small G12D cohort. These findings are exploratory and hypothesis-generating,
requiring prospective validation in larger, allele-resolved cohorts. They nonetheless
support continued investigation of allele-specific biology and dedicated therapeutic
development for KRAS G12D.


\end{abstract}

\begin{figure}[ht!]
\centering
\includegraphics[width=0.7\textwidth]{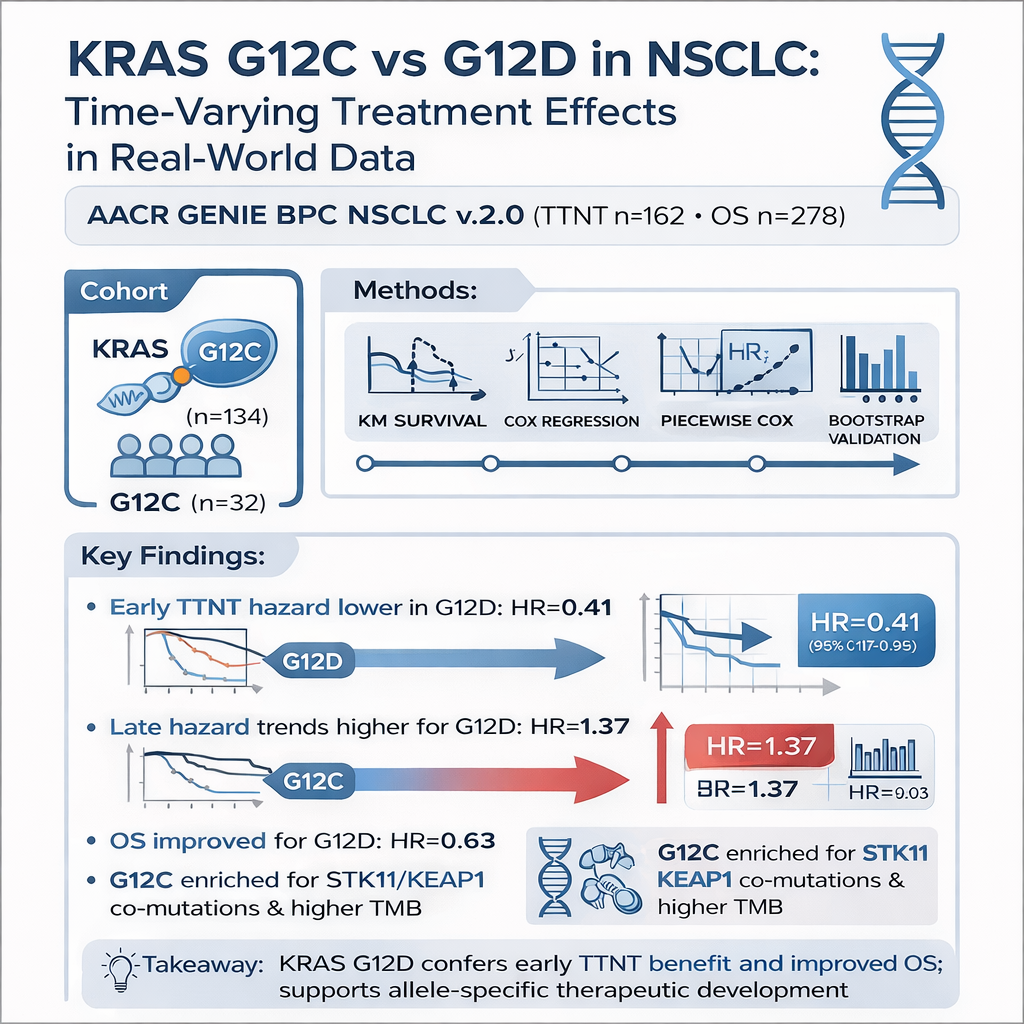}
\caption{Graphical result }
\label{}
\end{figure}

\newpage

% ============================================================
\section{Introduction}

KRAS mutations are the most prevalent oncogenic drivers in non--small cell lung
cancer (NSCLC), occurring in approximately 25--30\% of lung adenocarcinomas
worldwide~\citep{Frisch2025KRASReview, LungCancer2023KRASPrevalence}.
Among these, the glycine-to-cysteine substitution (G12C) accounts for approximately
13\% of all NSCLC cases, while codon-12 variants including G12D, G12V, and G12S
contribute to the remaining molecular heterogeneity~\citep{RWE2025G12CG12D}.
This allele-level heterogeneity is clinically meaningful: KRAS substitutions differ
in downstream biochemical signaling, co-mutation profiles, tumor immune
microenvironment (TIME) characteristics, and sensitivity to therapeutic
agents~\citep{Frontiers2025Comutations, AnnOnc2023Biomarkers}.

The development of allele-specific KRAS G12C inhibitors has transformed treatment
options for this molecular subgroup. Sotorasib and adagrasib have produced
clinically meaningful response rates and progression-free survival benefits in
previously treated KRAS G12C NSCLC~\citep{Sotorasib2021, Adagrasib2022}, and
next-generation agents such as divarasib continue to expand the landscape~\citep{Divarasib2024}.
By contrast, no approved targeted therapies exist for KRAS G12D, although
early-phase G12D-directed agents including zoldonrasib have shown preliminary
activity~\citep{Zoldonrasib2024}. These divergent therapeutic trajectories
underscore the importance of characterizing allele-specific clinical behavior
in real-world populations.

Real-world evidence suggests that allele-specific outcomes may be substantially
modified by co-occurring genomic alterations. KRAS G12C tumors frequently exhibit
higher tumor mutational burden (TMB), tobacco-related mutational signatures, and
enriched baseline immunogenicity relative to non-G12C alleles~\citep{MetaAnalysis2023KRASICI, JClinMed2025KRASICI}.
Concurrently, alterations in \textit{STK11}, \textit{KEAP1}, and \textit{SMARCA4}
define immunologically cold phenotypes with attenuated benefit from immune checkpoint
blockade (ICB)~\citep{AnnOnc2023Biomarkers}, and these co-mutations occur at
differing frequencies across KRAS alleles~\citep{CancerCell2025ATRCHK1,
Frontiers2025Comutations}.

Most existing retrospective studies of KRAS allele outcomes rely on static survival
comparisons or trial-enriched cohorts, limiting real-world generalizability~\citep{RWE2025G12CG12D}.
Importantly, conventional Cox proportional-hazards models assume a constant hazard
ratio over the entire follow-up period; if allele-specific effects are time-varying,
such analyses may obscure clinically meaningful dynamics. Time-to-next-treatment
(TTNT) has emerged as a pragmatic and validated real-world surrogate for
progression-free survival in observational oncology datasets, permitting
longitudinal assessment of treatment durability~\citep{GENIEBPC2022,
Khozin2019RWD, Garrido2022RWE_TTNT}.

The AACR Project GENIE BioPharma Collaborative (BPC) NSCLC dataset provides a
uniquely comprehensive platform for allele-resolved, time-adaptive outcome analyses,
integrating harmonized multi-institutional clinical timelines with matched genomic
profiles~\citep{GENIE2023Cohort, GENIEBPC2022}. Using this resource, we sought to:
(1) compare TTNT and OS between KRAS G12C and G12D in a real-world multi-institutional
cohort; (2) evaluate whether allele-specific hazards are time-varying using
piecewise and time-dependent Cox models; and (3) characterize allele-specific
co-mutation profiles and assess their relationship to outcomes.

% ============================================================
\section{Methods}

\subsection{Data Source}
De-identified clinical and genomic data were obtained from the AACR Project GENIE
BioPharma Collaborative (BPC) NSCLC v2.0-public release, accessed through
cBioPortal~\citep{GENIEBPC2022, AACRGENIE2023}. This dataset includes harmonized
clinical annotations, curated systemic treatment timelines, and linked somatic
mutation, copy-number alteration, and structural variant files for patients treated
at multiple U.S. academic cancer centers. Because all data are fully de-identified,
this analysis was exempt from institutional review board oversight.

\subsection{Cohort Construction}
\begin{figure}[ht]
\centering
\begin{tikzpicture}[
  node distance=9mm,
  every node/.style={
    rectangle, draw, rounded corners,
    align=center, font=\small,
    text width=8cm, minimum height=10mm
  },
  arrow/.style={-stealth, thick}
]
\node (all)  {\textbf{1,775 NSCLC patients} \\ AACR GENIE BPC v2.0-public};
\node (kras) [below=of all]
  {\textbf{KRAS-mutant subset} \\ G12C = 214;\quad G12D = 64};
\node (elig) [below=of kras]
  {Evaluable treatment timelines,\\ TTNT months and event status};
\node (anal) [below=of elig]
  {\textbf{TTNT analytic cohort}\\ $n=162$ (G12C=130; G12D=32)};
\node (splt) [below=of anal]
  {\textbf{Piecewise Cox dataset}\\ 243 start--stop intervals (162 patients)};
\node (os)   [right=38mm of anal]
  {\textbf{OS analytic cohort}\\ $n=278$ (133 deaths)};
\draw[arrow] (all)  -- (kras);
\draw[arrow] (kras) -- (elig);
\draw[arrow] (elig) -- (anal);
\draw[arrow] (anal) -- (splt);
\draw[arrow] (anal.east) -- ++(1.5,0) |- (os.west);
\end{tikzpicture}
\caption{CONSORT-style flow diagram of patient selection for TTNT and OS analyses.}
\label{fig:flow}
\end{figure}
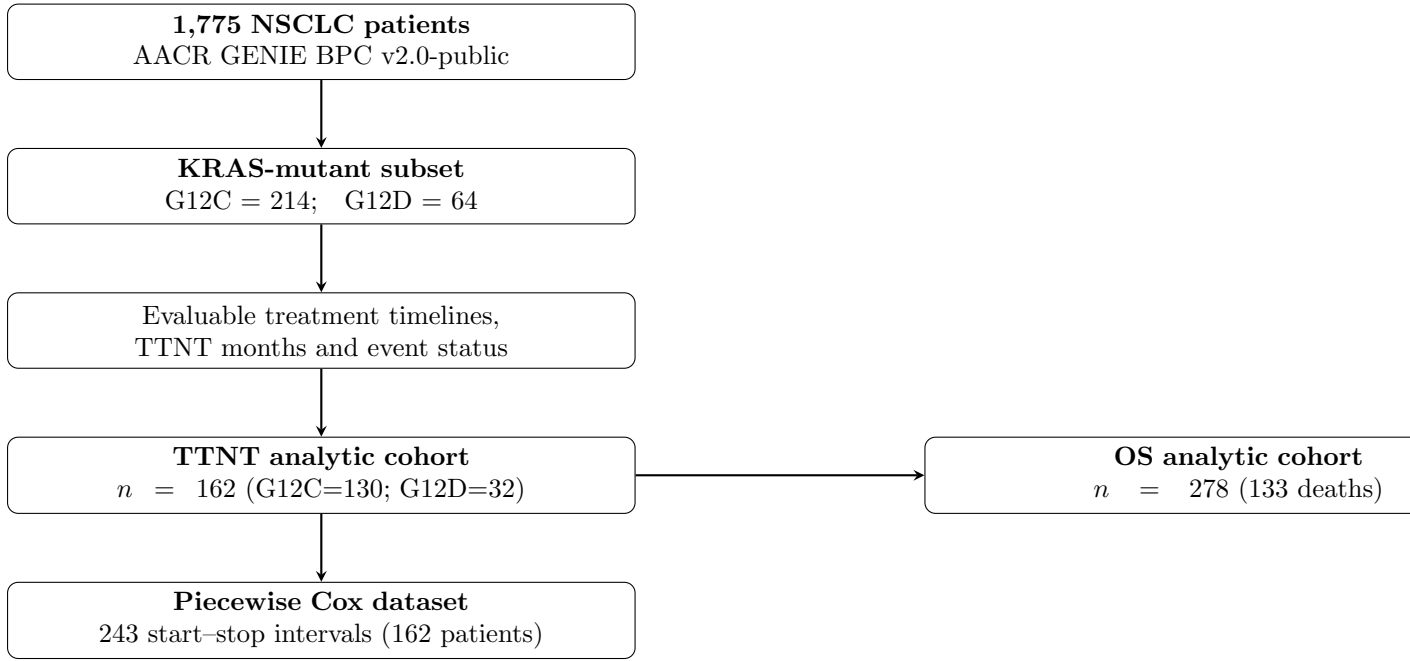

Patients were included if they had: (1) a diagnosis of lung adenocarcinoma;
(2) a confirmed somatic KRAS p.G12C or p.G12D mutation on next-generation
sequencing; and (3) evaluable systemic therapy timelines enabling derivation of
TTNT. Patients lacking therapy start dates, missing TTNT event indicators, or
without linked genomic data were excluded. OS analyses used available
diagnosis-to-death timelines and harmonized vital status fields.
Cohort construction is summarized in Figure~\ref{fig:flow}.

\subsection{Endpoint Definitions}

\paragraph{Time-to-Next-Treatment (TTNT).}
TTNT served as the primary real-world endpoint and pragmatic surrogate for
progression-free survival~\citep{Khozin2019RWD, Garrido2022RWE_TTNT}.
TTNT was defined as the maximum recorded duration across all systemic therapy
episodes in GENIE BPC. The event indicator was the initiation of a subsequent
regimen (number of cancer-directed regimens $> 1$); patients on a single regimen
were censored at last follow-up. This definition reflects clinician-assessed
treatment failure or progression necessitating a regimen change.

\paragraph{Overall Survival (OS).}
OS was measured from the date of pathological diagnosis to death or last known
contact. Vital status was derived from harmonized ``OS Status from start of
[therapy]'' fields across all curated treatment lines; a patient was classified
as deceased if any field indicated ``1:DECEASED o.''

\subsection{Genomic Variable Harmonization}
Binary alteration indicators for five co-mutation genes of clinical interest
(\textit{TP53}, \textit{STK11}, \textit{KEAP1}, \textit{SMARCA4}, \textit{MET})
were derived by merging mutation, copy-number alteration, and structural variant
tables. A gene was considered altered if any data source indicated a non-wild-type
call; samples with missing data were conservatively assigned wild-type status.
TMB was extracted from the curated nonsynonymous mutation count field and analyzed
as a continuous variable. PD-L1 positivity was binarized (positive vs.
negative/unknown) using harmonized institutional report values.

\subsection{Statistical Analysis}

\subsubsection{Descriptive and Comparative Analyses}
Baseline characteristics were summarized by allele group using medians for
continuous variables and proportions for categorical variables.
Between-group differences were assessed with Wilcoxon rank-sum tests (continuous)
and Fisher's exact tests (categorical).

\subsubsection{Kaplan--Meier and Standard Cox Models}
Unadjusted TTNT and OS were compared using Kaplan--Meier curves and log-rank tests.
Multivariable Cox proportional-hazards models were adjusted for age at sequencing,
sex, and \textit{STK11} and \textit{TP53} co-mutation status. Proportional hazards
assumptions were assessed using Schoenfeld residuals for each covariate and globally.

\subsubsection{Time-Varying Hazard Models}
Because KRAS exhibited borderline non-proportionality (Schoenfeld $p = 0.053$),
two complementary time-adaptive approaches were applied:

\paragraph{Time-dependent Cox model.}
A KRAS $\times \log(t)$ interaction term was fitted to quantify continuous temporal
change in the allele-specific hazard, without imposing a fixed cut-point.

\paragraph{Pre-specified piecewise Cox model.}
Follow-up time was partitioned at the median TTNT (23.0 months), a cut-point
selected \textit{a priori} based on the distribution of follow-up times rather than
outcome-dependent optimization, thereby avoiding data-driven selection bias.
Using the counting-process data structure, each patient contributed one or two
start--stop intervals ($[0,c]$ and $(c, T_i]$). The KRAS$\times$Period interaction
term quantified the change in allele-specific hazard between early and late intervals.
Period-specific HRs and 95\% confidence intervals were derived using the
delta method.

\subsubsection{Bootstrap Internal Validation}
To assess the robustness of early and late HR estimates, nonparametric bootstrap
resampling was performed ($B = 1000$ iterations). In each iteration, patients were
resampled with replacement at the patient level, the piecewise Cox model was refit,
and period-specific HRs were extracted. Results were summarized as medians and
95\% percentile confidence intervals across bootstrap replicates.

\subsubsection{Sensitivity Analyses}
Pre-specified sensitivity analyses included: (1) adjustment for TMB (per 5 mut/Mb)
to evaluate potential confounding by mutational burden; (2) alternative piecewise
cut-points at the 25th, 75th, and 95th TTNT percentiles to assess cut-point
dependence; and (3) exclusion of treatment changes within 2 months to reduce
administrative censoring bias.

All analyses were performed in R (version~4.3.1) using the \texttt{survival},
\texttt{survminer}, \texttt{flexsurv}, \texttt{dplyr}, and \texttt{ggplot2}
packages~\citep{Therneau2023Survival, Kassambara2019Survminer, Wickham2023Dplyr}.

%\newpage

\begin{figure}[ht!]
\centering
\includegraphics[width=0.7\textwidth]{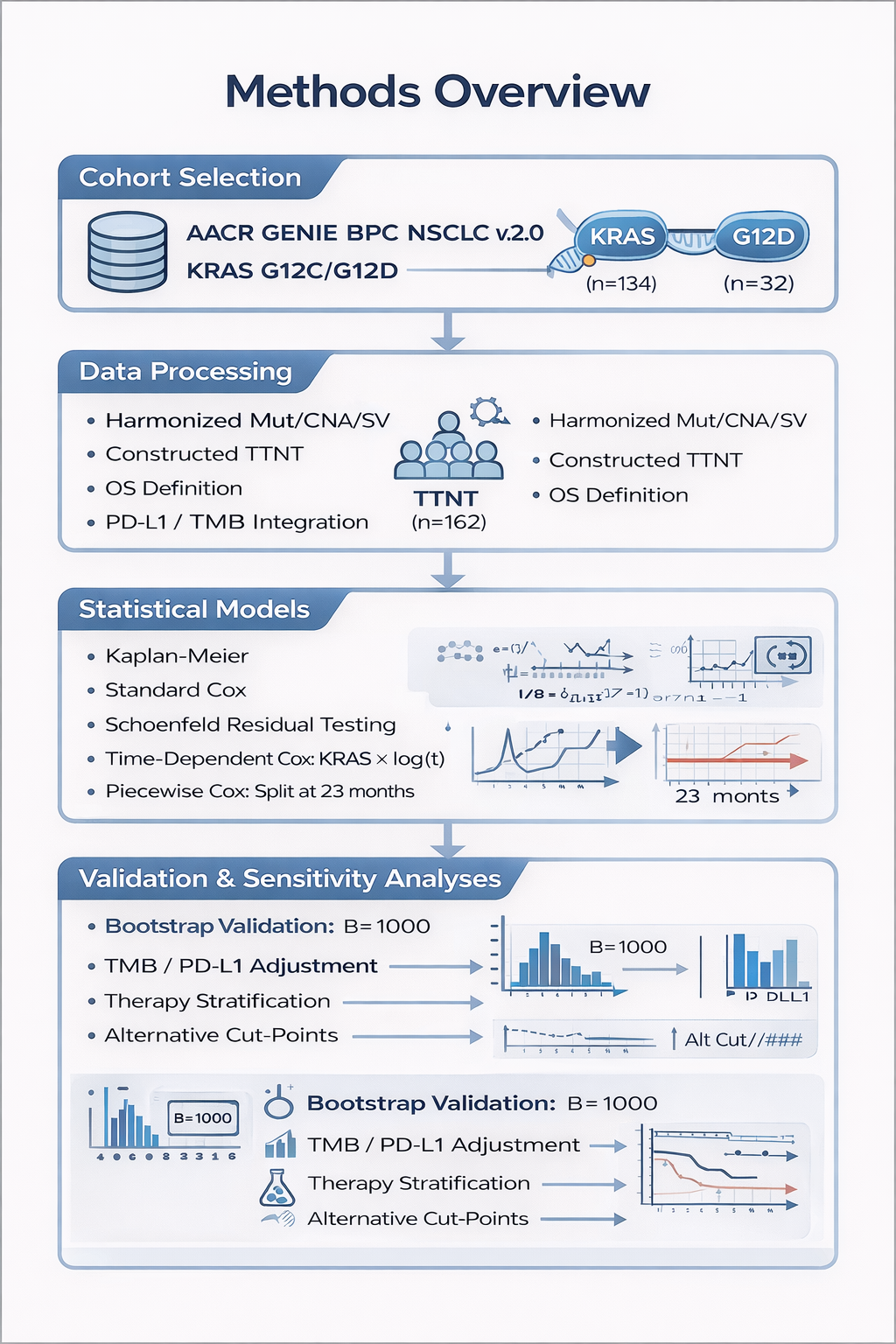}
\caption{}
\label{}
\end{figure}

% ============================================================

\newpage

\section{Results}

\subsection{Cohort Characteristics and Molecular Features}

Of 1,775 NSCLC patients in the AACR GENIE BPC v2.0-public dataset, 278 patients
with KRAS G12C ($n = 214$) or KRAS G12D ($n = 64$) mutations had evaluable OS data
and formed the molecular characterization cohort. A subset of 162 patients met all
criteria for TTNT analysis (G12C $n = 130$; G12D $n = 32$; Table~\ref{tab:baseline}).

KRAS G12C tumors showed significantly higher median TMB (9.79 vs.\ 7.83 mut/Mb;
Wilcoxon $p = 0.002$) and greater co-mutation enrichment in \textit{STK11}
(31.8\% vs.\ 17.2\%; Fisher's $p = 0.027$). \textit{KEAP1} alteration showed a
similar directional trend (18.7\% vs.\ 9.4\%; $p = 0.087$). \textit{TP53},
\textit{SMARCA4}, and \textit{MET} frequencies did not differ significantly between
alleles. PD-L1 positivity rates were comparable (57\% G12C vs.\ 50\% G12D;
$p = 0.623$).

\begin{table}[ht]
\centering
\caption{Baseline clinical and genomic characteristics by KRAS allele (OS cohort,
$n = 278$). The TTNT cohort ($n = 162$) is a subset with complete treatment
timeline data. TMB = tumor mutational burden; PD-L1 = programmed death-ligand~1.
Continuous variables compared by Wilcoxon rank-sum test; categorical variables
by Fisher's exact test.}
\begin{tabular}{lcccc}
\toprule
\textbf{Characteristic}
  & \textbf{G12C ($n=214$)}
  & \textbf{G12D ($n=64$)}
  & \textbf{Total ($N=278$)}
  & \textbf{$p$-value} \\
\midrule
Sex: male, $n$ (\%)      & 82 (38.3\%) & 19 (29.7\%) & 101 (36.3\%) & 0.238 \\
TMB, median (mut/Mb)     & 9.79        & 7.83        & 9.40         & 0.002 \\
PD-L1 positive (\%)      & 49 (57.0\%) & 10 (50.0\%) & 59 (55.1\%)  & 0.623 \\
any-\textit{STK11} (\%)  & 68 (31.8\%) & 11 (17.2\%) & 79 (28.4\%)  & 0.027 \\
any-\textit{TP53} (\%)   & 92 (43.0\%) & 27 (42.2\%) & 119 (42.8\%) & 1.000 \\
any-\textit{KEAP1} (\%)  & 40 (18.7\%) &  6 (9.4\%)  & 46 (16.5\%)  & 0.087 \\
any-\textit{SMARCA4} (\%)& 27 (12.6\%) &  5 (7.8\%)  & 32 (11.5\%)  & 0.375 \\
any-\textit{MET} (\%)    & 21 (9.8\%)  &  3 (4.7\%)  & 24 (8.6\%)   & 0.309 \\
\bottomrule
\end{tabular}
\label{tab:baseline}
\end{table}

\subsection{TTNT: Kaplan--Meier and Multivariable Cox Analyses}

\begin{figure}[ht!]
\centering
\includegraphics[width=0.7\textwidth]{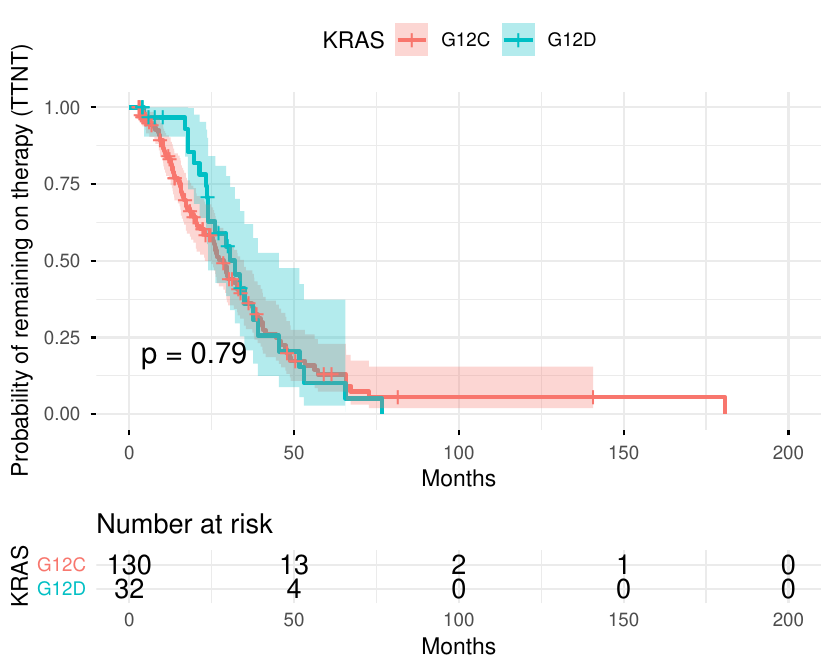}
\caption{Kaplan--Meier curves for time-to-next-treatment (TTNT) by KRAS allele
(G12C $n=130$; G12D $n=32$). Median TTNT: 28.6 vs.\ 32.0 months;
log-rank $p = 0.787$.}
\label{fig:KM_TTNT}
\end{figure}

Kaplan--Meier analysis demonstrated no significant overall TTNT difference between
alleles (Figure~\ref{fig:KM_TTNT}). Median TTNT was 28.6 months (95\%~CI: 24.9--34.7)
for G12C versus 32.0 months (95\%~CI: 23.9--45.4) for G12D (log-rank $p = 0.787$).

In multivariable Cox regression adjusting for age, sex, \textit{STK11}, and
\textit{TP53}, KRAS allele subtype was not significantly associated with overall
TTNT hazard (HR$_{\text{G12D vs G12C}} = 0.85$; 95\%~CI 0.53--1.37; $p = 0.50$;
Table~\ref{tab:cox}). \textit{STK11} alteration showed a directional trend toward
shorter TTNT (HR = 1.30; 95\%~CI 0.85--1.99; $p = 0.23$) consistent with published
literature, though it did not reach statistical significance in this sample.

Schoenfeld residual testing indicated borderline non-proportionality for the KRAS
term ($p_{\text{KRAS}} = 0.053$; global $p = 0.334$), motivating additional
time-adaptive modeling.

\begin{table}[ht]
\centering
\caption{Multivariable Cox proportional-hazards model for TTNT.}
\label{tab:cox}
\begin{tabular}{lccc}
\toprule
\textbf{Variable} & \textbf{HR} & \textbf{95\%~CI} & \textbf{$p$-value} \\
\midrule
KRAS G12D vs G12C & 0.85 & 0.53--1.37 & 0.50 \\
Age (per year)    & 1.01 & 0.99--1.03 & 0.55 \\
Sex: Female (ref: Male) & 1.35 & 0.87--2.10 & 0.18 \\
\textit{STK11} altered  & 1.30 & 0.85--1.99 & 0.23 \\
\textit{TP53} altered   & 0.77 & 0.52--1.15 & 0.20 \\
\bottomrule
\end{tabular}
\end{table}

\subsection{Time-Varying Hazard Analysis}

\subsubsection{Piecewise Cox Model}

\begin{figure}[ht!]
\centering
\includegraphics[width=0.65\textwidth]{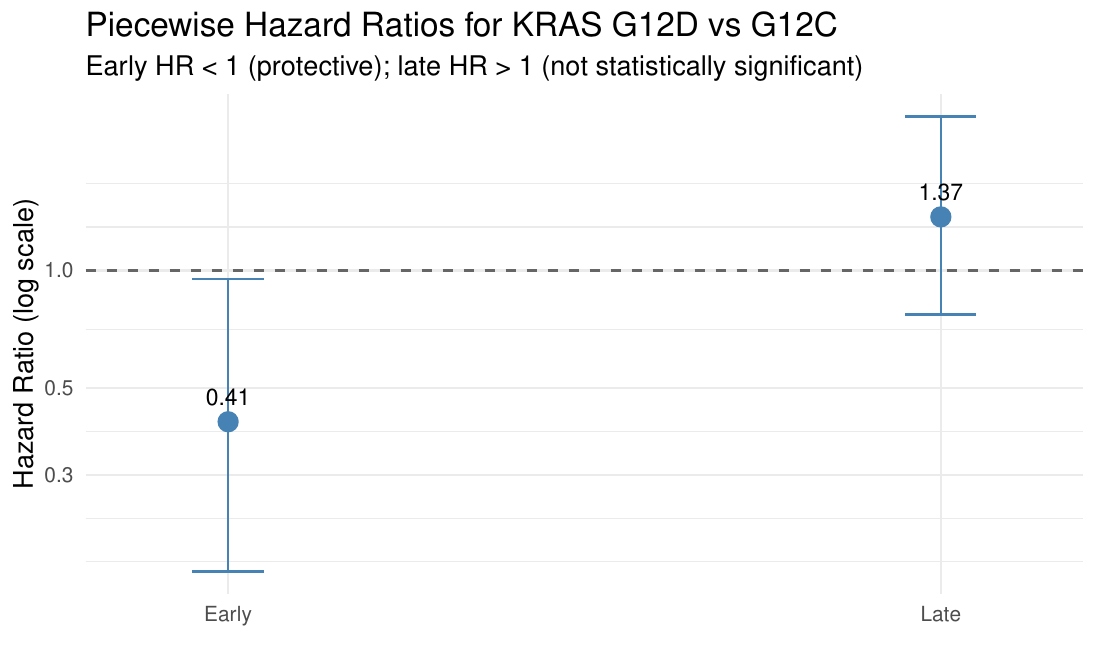}
\caption{Period-specific hazard ratios for KRAS G12D vs G12C from the piecewise
Cox model (cut-point: 23 months). HR$_{\text{early}} = 0.41$ (95\%~CI 0.17--0.97);
HR$_{\text{late}} = 1.38$ (95\%~CI 0.77--2.47). Dashed line at HR = 1.0.
Late-period HR did not reach statistical significance ($p = 0.285$).}
\label{fig:piecewise_HR}
\end{figure}

The pre-specified piecewise Cox model (cut at median TTNT = 23.0 months) generated
243 start--stop intervals from 162 patients contributing 114 events.
The model revealed a time-varying hazard pattern
(Table~\ref{tab:piecewise_cox}; Figure~\ref{fig:piecewise_HR}).

During the early interval (0--23 months), G12D was associated with a lower TTNT
hazard compared with G12C (HR$_{\text{early}} = 0.41$; 95\%~CI 0.17--0.97;
$p = 0.043$). A statistically significant KRAS$\times$Period interaction
(HR = 3.33; 95\%~CI 1.19--9.31; $p = 0.021$) indicated that this early difference
did not persist: the late-period HR was 1.38 (95\%~CI 0.77--2.47; $p = 0.285$),
with the confidence interval crossing unity and not reaching statistical significance.
The interaction indicates a directional shift in the relative hazard between periods,
though late-period estimates should be interpreted with caution given their
imprecision (see Limitations). Overall model fit was supported by concordance = 0.614
(SE = 0.029) and a significant likelihood ratio test ($p = 0.04$).

Complementary time-dependent Cox modeling using a KRAS$\times\log(t)$ interaction
produced directionally consistent findings, confirming that the temporal pattern is
not an artifact of the specific 23-month cut-point choice.

\begin{table}[ht!]
\centering
\caption{Piecewise Cox model for TTNT. The KRAS coefficient represents the
early-period HR; the late-period HR (1.38; 95\%~CI 0.77--2.47) is derived from
the sum of the KRAS and interaction coefficients using the delta method.}
\begin{tabular}{lccc}
\toprule
\textbf{Variable} & \textbf{HR} & \textbf{95\%~CI} & \textbf{$p$-value} \\
\midrule
KRAS G12D (early period)           & 0.41 & 0.17--0.97 & 0.043 \\
Age (per year)                     & 1.01 & 0.99--1.03 & 0.532 \\
Sex (Male vs Female)               & 0.73 & 0.47--1.12 & 0.149 \\
\textit{STK11} altered             & 1.22 & 0.79--1.89 & 0.359 \\
\textit{TP53} altered              & 0.78 & 0.52--1.17 & 0.225 \\
KRAS $\times$ late period          & 3.33 & 1.19--9.31 & 0.021 \\
\midrule
\multicolumn{4}{l}{\small $n = 162$ patients; events = 114;
  concordance = 0.614 (SE = 0.029)} \\
\multicolumn{4}{l}{\small Likelihood ratio $p = 0.04$;
  Wald $p = 0.05$; Score $p = 0.05$} \\
\multicolumn{4}{l}{\small Late-period KRAS HR (delta method):
  1.38 (95\%~CI 0.77--2.47; $p = 0.285$)} \\
\bottomrule
\end{tabular}
\label{tab:piecewise_cox}
\end{table}

\subsubsection{Bootstrap Validation}

\begin{figure}[ht!]
\centering
\includegraphics[width=0.70\textwidth]{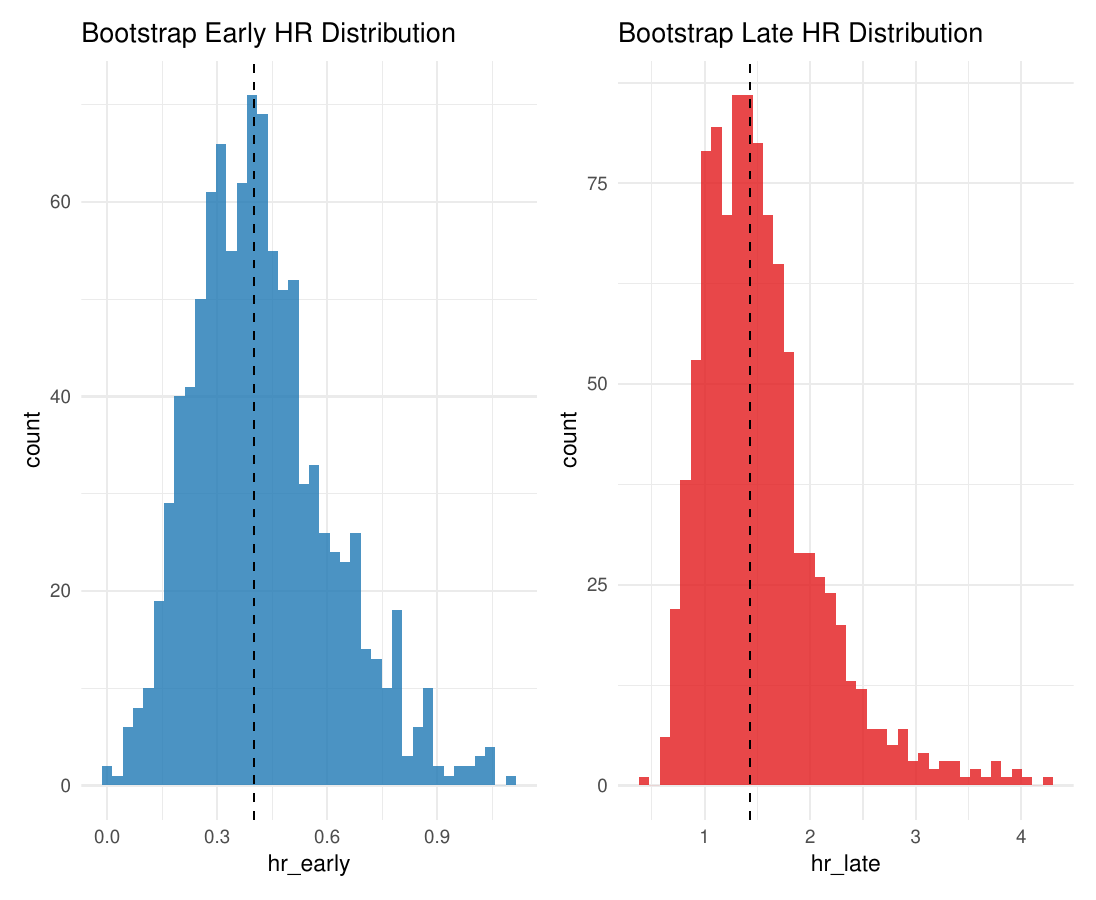}
\caption{Bootstrap distributions ($B = 1000$ resamples) for early and late
period-specific hazard ratios. Median HR$_{\text{early}} = 0.39$
(95\% PI: 0.11--0.87); median HR$_{\text{late}} = 1.41$ (95\% PI: 0.75--3.24).
Dashed vertical lines indicate medians; solid red lines indicate HR = 1.}
\label{fig:bootstrap_HR}
\end{figure}

Nonparametric bootstrap resampling ($B = 1000$) at the patient level confirmed the
directional stability of the time-varying pattern (Figure~\ref{fig:bootstrap_HR}).
Bootstrap-derived HR distributions yielded median HR$_{\text{early}} = 0.39$
(95\% percentile interval: 0.11--0.87) and median HR$_{\text{late}} = 1.41$
(95\% percentile interval: 0.75--3.24). The close agreement between point estimates
and bootstrap medians supports the internal robustness of the observed pattern.
The wider percentile interval for the late HR reflects genuine uncertainty arising
from the limited G12D sample size (see Limitations).

\subsection{Overall Survival}

\begin{figure}[ht!]
\centering
\includegraphics[width=0.70\textwidth]{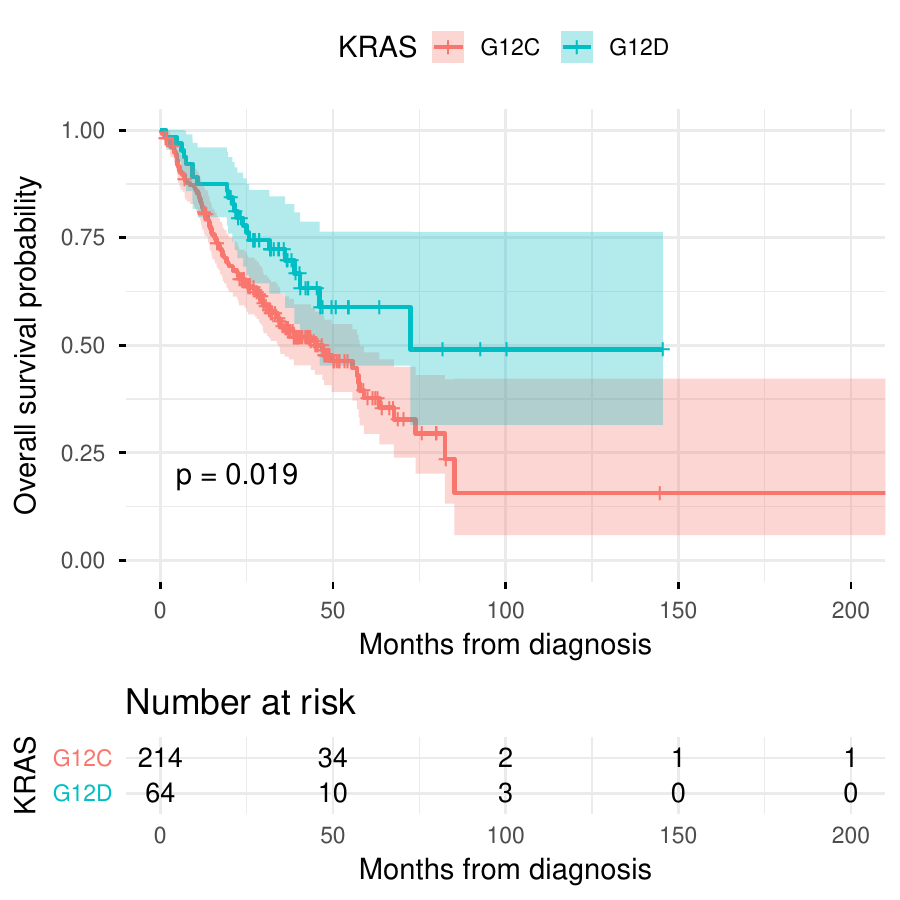}
\caption{Kaplan--Meier overall survival curves by KRAS allele (months from
diagnosis; $n = 278$; 133 deaths). Log-rank $p = 0.019$ (unadjusted). Adjusted
HR for G12D = 0.63 (95\%~CI 0.39--0.99; $p = 0.048$; Table~\ref{tab:cox_os}).}
\label{fig:os_km}
\end{figure}

Among 278 OS-evaluable patients (G12C $n = 214$; G12D $n = 64$; 133 deaths),
Kaplan--Meier analysis showed improved survival for G12D (log-rank $p = 0.019$;
Figure~\ref{fig:os_km}). Median OS was 44.8 months for G12C versus 72.4 months
for G12D. After multivariable adjustment for age, sex, \textit{STK11}, and
\textit{TP53}, the OS benefit for G12D remained statistically significant
(HR = 0.63; 95\%~CI 0.39--0.99; $p = 0.048$; Table~\ref{tab:cox_os}).
Proportional hazards assumptions were satisfied (global $p = 0.91$). Age
(HR = 1.02 per year; $p = 0.023$) and \textit{STK11} alteration
(HR = 1.48; $p = 0.042$) were additional significant predictors.

\newpage

\begin{table}[ht!]
\centering
\caption{Multivariable Cox proportional-hazards model for overall survival (OS).}
\label{tab:cox_os}
\begin{tabular}{lcccc}
\toprule
\textbf{Variable} & \textbf{HR} & \textbf{95\%~CI} & \textbf{z} & \textbf{$p$-value} \\
\midrule
KRAS G12D vs G12C & 0.63 & 0.39--0.99 & $-1.98$ & 0.048 \\
Age (per year)    & 1.02 & 1.00--1.04 &  $2.27$ & 0.023 \\
Sex: Male         & 1.01 & 0.70--1.46 &  $0.05$ & 0.960 \\
\textit{STK11} altered & 1.48 & 1.01--2.16 & $2.03$ & 0.042 \\
\textit{TP53} altered  & 1.41 & 0.99--2.01 & $1.91$ & 0.056 \\
\bottomrule
\end{tabular}
\end{table}

\subsection{Sensitivity and Robustness Analyses}

\paragraph{TMB-adjusted model.}
Addition of TMB (per 5 mut/Mb) as a covariate did not materially alter the allele
effect: HR$_{\text{early}} = 0.39$ (95\%~CI 0.16--0.92); interaction HR = 3.38
(95\%~CI 1.21--9.44; $p = 0.020$). TMB itself was not significantly associated
with TTNT in the piecewise model ($p = 0.319$).

\paragraph{Alternative cut-points.}
Piecewise models at the 25th (12.0 months), 75th (34.7 months), and 95th
(65.4 months) TTNT percentiles consistently yielded early HRs below 1.0
(range: 0.19--0.83) and late HRs above 1.0 (range: 1.02--1.42), supporting
directional consistency across cut-point choices, though confidence intervals
widened with more extreme splits due to smaller risk sets.

\paragraph{Exclusion of early events ($\le$2 months).}
Censoring treatment changes within 2 months did not alter the primary findings:
HR$_{\text{early}} = 0.41$ (95\%~CI 0.18--0.97); interaction HR = 3.33
(95\%~CI 1.19--9.31; $p = 0.022$).

% ============================================================
\section{Discussion}

In this real-world, allele-resolved analysis of the AACR Project GENIE BPC NSCLC
cohort, we found that KRAS G12C and G12D lung adenocarcinomas exhibit distinct
co-mutation profiles and directionally different temporal treatment durability
patterns. Although unadjusted TTNT and standard multivariable Cox models did not
identify a significant overall difference between alleles, time-structured modeling
revealed a statistically supported early TTNT advantage for G12D that attenuated
during later follow-up. This time-varying pattern was directionally consistent
across bootstrap resampling, alternative cut-points, and TMB-adjusted sensitivity
analyses. Additionally, G12D was associated with significantly improved OS in a
larger evaluable cohort.

\subsection{Biological Context for Observed Patterns}

The allele-specific molecular features observed in our cohort provide a plausible
biological framework for the time-varying outcomes. G12C tumors displayed
significantly higher TMB and greater enrichment of \textit{STK11} and \textit{KEAP1}
co-mutations, findings concordant with prior translational and registry
studies~\citep{Skoulidis2021, Prior2020, Frontiers2025Comutations}. \textit{STK11}
and \textit{KEAP1} alterations define immunologically cold tumor phenotypes with
attenuated benefit from ICB and are associated with poorer outcomes across KRAS
alleles~\citep{Skoulidis2021, AnnOnc2023Biomarkers}. The relative paucity of these
co-mutations in G12D tumors may contribute to more durable early treatment responses,
including to ICB-containing regimens commonly used in first-line NSCLC.

The G12D OS advantage (adjusted HR = 0.63; $p = 0.048$) persisted despite G12D's
lower median TMB and late-period TTNT convergence. This is noteworthy because lower
TMB is generally associated with reduced immunotherapy benefit, suggesting that the
co-mutation landscape — particularly the relative absence of \textit{STK11}/\textit{KEAP1}
alterations — may preserve treatment responsiveness across sequential lines in G12D
tumors more than TMB alone would predict. These observations are concordant with
published evidence that co-mutation context, rather than TMB in isolation, is a
more reliable predictor of KRAS allele-specific ICB outcomes~\citep{MetaAnalysis2023KRASICI,
ClinLungCancer2025ICIComutations}.

The late-period attenuation of G12D's early TTNT advantage may reflect several
non-mutually exclusive mechanisms: emergence of resistant subclones, clonal evolution
under therapeutic pressure, exhaustion of immune-mediated tumor control, or
differential downstream KRAS signaling over time. These dynamic changes would not
be detectable using conventional proportional-hazards
models~\citep{SkoulidisHeymach2020, MechanismsKRAS2025}, underscoring the value of
time-adaptive analytical frameworks for molecular subgroup analyses.

\subsection{Therapeutic Context}

An important consideration is that G12C-targeted inhibitors (sotorasib, adagrasib)
became clinically available during the study period covered by the GENIE BPC dataset.
Our analysis could not separately identify patients who received these agents.
If G12C patients disproportionately accessed targeted therapies in later lines, this
could contribute to the attenuation of G12D's early durability advantage independent
of biological mechanisms. Future analyses with explicit capture of allele-targeted
therapy exposure will be necessary to separate treatment-era effects from underlying
allele biology.

For G12D specifically, early-phase inhibitors including zoldonrasib represent the
first targeted options for this subgroup~\citep{Zoldonrasib2024}. Our real-world
TTNT and OS data provide pre-targeted-era benchmarks that may contextualize future
comparative effectiveness analyses as G12D-directed therapies mature.

\subsection{Limitations}

This study has several important limitations that require careful consideration.

First and most critically, the G12D cohort in the TTNT analyses was small ($n = 32$),
which substantially constrained statistical power, particularly for late-period
inferences. Post-hoc power analysis using the Schoenfeld formula indicated that
detecting the observed late-period HR of 1.38 with 80\% power at $\alpha = 0.05$
would require approximately 489 events among patients surviving beyond 23 months,
corresponding to roughly 193 G12D and 785 G12C patients. With approximately 40
late-period events observed, the achieved power for the late-period comparison was
only 12.3\%. Consequently, the non-significant late-period HR ($p = 0.285$) most
plausibly reflects inadequate power rather than absence of a true effect, and
late-period estimates should be treated as exploratory and hypothesis-generating.
Confirmation of the late-period pattern requires prospective validation in larger,
allele-resolved cohorts or pooled multi-institutional analyses.

Second, TTNT is an imperfect surrogate for radiographic progression: clinician
decisions to change therapy may be influenced by tolerability, patient preference,
or administrative factors independent of disease progression. Nonetheless, TTNT
is a widely used and pragmatically validated real-world endpoint in observational
oncology research~\citep{Khozin2019RWD, Garrido2022RWE_TTNT}. Third, incomplete
or variable reporting of genomic data (CNAs, SVs, TMB, PD-L1) across institutions
may introduce differential misclassification of co-mutation and biomarker variables.
Fourth, unmeasured confounding by treatment selection, center-specific practice
patterns, or performance status not captured in the BPC public release  cannot
be fully excluded despite multivariable adjustment. Fifth, OS follow-up in the
BPC v2.0 public release may be incomplete or variably mature across centers, and
the OS findings, while statistically significant, should be interpreted in that context.

\subsection{Strengths}

Strengths include the use of deeply curated, harmonized treatment timelines from
GENIE BPC enabling high-resolution TTNT measurement; comprehensive co-mutation
harmonization across three genomic data types; application of multiple
complementary survival modeling frameworks that explicitly evaluated proportional
hazards assumptions; and rigorous bootstrap internal validation demonstrating
directional stability of time-varying estimates. Pre-specifying the median
cut-point prior to analysis avoids data-driven selection bias that commonly
inflates type-I error in piecewise modeling.

\subsection{Implications and Future Directions}

These findings support several practical implications. First, KRAS-mutant NSCLC
should not be treated as a uniform molecular entity in clinical or research settings;
allele-specific and temporally stratified outcome reporting is warranted. Second,
prospective studies and registries should prespecify time-varying survival analyses
and ensure sufficient allele-level sample sizes to detect dynamic effects with
adequate power. Third, integrative multi-omic approaches—including immune
profiling, longitudinal circulating tumor DNA sampling, and post-progression genomic
characterization, are needed to mechanistically dissect the early-to-late transition
in allele-specific hazards. Finally, patients with KRAS G12D should be considered
high-priority candidates for G12D-directed clinical trials as targeted agents
advance through development.

\subsection{Conclusions}

In this multi-institutional real-world analysis, KRAS G12D lung adenocarcinomas
demonstrated an early TTNT durability advantage and significantly improved OS
compared with G12C, despite similar overall TTNT. Time-structured survival modeling
identified allele-specific hazard dynamics that conventional proportional-hazards
analyses did not reveal. The early-period KRAS effect was statistically significant
and bootstrap-confirmed; the late-period pattern was directionally consistent but
did not reach significance, reflecting limited power from the small G12D cohort
rather than evidence against the effect. These exploratory findings challenge
the treatment of KRAS-mutant NSCLC as a homogeneous entity and support prospective
investigation of allele-specific treatment trajectories, particularly as G12D-targeted
therapeutics enter clinical practice.

% ============================================================
\section*{Acknowledgements}
We thank the AACR Project GENIE Consortium and all contributing institutions for
public access to de-identified data.

\section*{Data Availability}
Data are publicly available from the AACR Project GENIE BPC NSCLC v2.0-public
release via cBioPortal.

\section*{Conflicts of Interest}
The authors declare no conflicts of interest.

\section*{Funding}
No specific funding was received for this work.

% ============================================================
\bibliographystyle{unsrtnat}
\bibliography{refs}

\end{document}